\documentclass[preprint,3p,times,twocolumn]{elsarticle}

\usepackage{graphicx}
\usepackage{amssymb}

\usepackage{float}
\usepackage{units}
\usepackage{textcomp}
\usepackage{slashed}
\usepackage{physics}
\usepackage{braket}
\usepackage{amsmath}
\usepackage{amsthm}
\usepackage{esint}
\usepackage[latin9, utf8]{inputenc}
\usepackage[none]{hyphenat}

\journal{Phys. Lett. B}

\begin{document}

\begin{frontmatter}


\title{An innovative approach for sketching the QCD phase diagram within the NJL model using Lagrange Multipliers }



\author{Angelo Mart\'{\i}nez and Alfredo Raya}

\address{Instituto de F\'{\i}sica y Matem\'aticas, Universidad Michoacana de San Nicol\'as de Hidalgo. Edificio C-3, Ciudad Universitaria. Francisco J. M\'ujica s/n, Col. Fel\'{\i}citas del R\'{\i}o, CP 58040, Morelia, Michoac\'an, Mexico.}

\begin{abstract}
We develop a new approach for sketching the quantum chromodynamics phase-diagram within the Nambu--Jona-Lasinio model for arbitrarily large values of the coupling constant, temperature and chemical potential based upon the strategy of Lagrange multipliers that constrains the corresponding gap equation and its mass gradient. Our approach distinguishes continuous from discountinuous phase transitions and thus is capable of locating the position of the Critical End Point from the thermodynamical parameters of the model alone, without direct reference to the chiral susceptibility or any other observable. The approach can be straightforwardly extended to those effective chiral quark models with a momentum independent constituent quark mass.
\end{abstract}


\end{frontmatter}


\section{Introduction}
It is widely accepted that after the Big-Bang, a number of phase transitions occurred as the fundamental soup expanded and cooled down.
Hadronic matter in particular undergone a transition from the so-called quark-gluon plasma to bound states of these entities, either mesons or baryons or other exotic states, only one microsecond after the primordial explosion~~\cite{BraunMunzinger:2008tz, Schwarz:2003du}. These early universe temperature and density conditions are met in relativistic heavy-ion collision experiments as the ones carried out at RHIC~\cite{Adams:2005dq} and LHC~\cite{ALICE:2017jyt}, where the dominant interactions taking place are precisely color interactions. Quantum Chromodynamics (QCD) is the theory of strong interactions among quarks and gluons and it is well established that at high energy, due to the property of asymptotic freedom QCD interactions are perturbative, and at low energy, the theory exhibits the phenomena of dynamical chiral symmetry breaking and confinement, both responsible for the nature of hadron spectra.  Nevertheless, QCD is  highly non-linear in this regime -precisely because the coupling becomes strong-, thus rendering senseless any attempt to describe these phenomena under perturbative arguments. Therefore, the detailed understanding of the chiral and confinement/deconfinement transitions seems hard to achieve.  To unveil the traits of these transitions, there have been recent experimental proposals, besides RHIC and LHC, in facilities like NICA~\cite{KEKELIDZE2017884, Kekelidze:2012zzb}, FAIR~\cite{Sturm2011, Frankfurt:2019uvk} to probe the behavior of hadronic matter in different regimes of temperature and baryon density.

On the theoretical side, non-perturbative  QCD can be described on a lattice~\cite{Sharma:2013hsa} or in continuum through field theoretical methods~\citep{Fischer:2018sdj, Alkofer:2000wg, Schaefer2008, PAWLOWSKI20072831, Reinhardt:2017pyr}. These first principles approaches allow to describe at zero density, the confinement/deconfinement and chiral symmetry breaking/restoration transitions so that we know they take place (roughly simultaneously) at a temperature $T=T_c\approx170-186\,{\rm MeV}$ for 2 light-quark flavors~\cite{Maezawa:2007fd} and up to $T=T_c\approx150-156.5\,{\rm MeV}$ when the strange quark is taken into account~\cite{PhysRevLett.113.082001, Bazavov:2011nk}. The transitions are a cross-over for finite physical-quark masses, and  correspond to a continuum second order transition in the chiral limit. On the other hand, effective models indicate that at zero temperature, but high density or large chemical potential $\mu$, the transition is discontinuous (first order). These two facts hint toward the existence of a Critical End Point (CEP) where the continuous and discontinuous phase transitions meet in the $T-\mu$ plane. Location of the CEP is at some extent the holy grail for experimental and theoretical efforts. On the former, fluctuations on the number of baryons and photons are linked to the correlation length in the phase transition and the susceptibilities~\cite{Stephanov:2011pb, Stephanov:2008qz, Alba:2014eba},  and thus are favorite signals to locate the CEP. For the later, the peaks of the thermal gradient of the chiral susceptibility~\cite{Xu_2015} and/or the Polyakov loop~\cite{Fukushima:2017csk} are selected for the location of the CEP. One should be aware that $T_c$ would change depending upon the observable and the corresponding susceptibilities under consideration.

In this letter we address the problem of locating the CEP in the $T\!-\!\mu$-plane within the Nambu-Jona-Lasinio (NJL) model~\cite{PhysRev.122.345, PhysRev.124.246}. Historically, the NJL model represents one of the first attempts to describe the interactions between nucleons, being fairly successful in the description of the dynamical  generation of nucleon mass through the phenomenon of spontaneous symmetry breaking, in analogy with superconductivity. The model is non-renormalizable, thus the regularization of the momentum integrals is a must. Nowadays, it has been promoted to a quark interaction model and given its simplicity, the incorporation of the medium effects is straightforward~\cite{Buballa:2003qv}. As the model was developed before the discovery of quarks and gluons, it is incapable of describing confinement. Nevertheless attempts had been made to incorporate confinement by coupling a Polyakov loop potential~\cite{Meisinger:1995ih, Fukushima:2003fw}, by regularizing integrals non-locally~\cite{GomezDumm:2006vz, General:2000zx} and by adopting a regularization procedure such that the quark propagator develops a pole-less structure~\citep{Ebert:1996vx, GutierrezGuerrero:2010md, Roberts:2007ji}. 

In our present study, we consider the quark gap equation and its derivative with respect to the mass (mass gradient) as a function of the temperature and chemical potential. 
Thus, with a strategy inspired in Lagrange multipliers, we identify the critical parameters $T_c$ and $\mu_c$ for which the gap equations supports none, one or more solutions, identifying the coupling as a Lagrange multiplier. A trivial solution $M=0$ is found in the chirally symmetric phase of the model, whereas a unique positive solution $M>0$ is found in a chirally broken phase through a continuous transition which in the chiral limit corresponds to a second order phase transition. When two positive solutions are found in the chirally broken phase,  we observe a discontinuous first order transition. Thus we can identify  straightforwardly the position of the CEP where the continuous and discontinuous phase transitions meet. To exemplify how our strategy works, we have organized the remainder of this letter as follows: In Sect.~\ref{S:1} we present the NJL model and its corresponding  gap equation in vacuum and in a medium. We then introduce the framework we present to analyze it and predict the position of the CEP in Sect.~\ref{S:2}. We finally discuss our findings and conclude in Sect.~\ref{S:3}.

\section{Gap equation of the NJL model}
\label{S:1}

The Lagrangian of the NJL model is~\cite{ PhysRev.122.345, PhysRev.124.246} 
\begin{equation}
    {\cal L}=\bar\psi \left( i {\not \! \partial} -m_q\right) \psi + G\left[ (\bar\psi\psi)^2+(\bar\psi i\gamma_5\vec\tau\psi)^2\right]\;.
\end{equation}
\noindent
Here, $\psi$ represents the quark field, $m_q$ is the current mass, $G$ is the coupling constant and $\vec{\tau}$ are the Pauli matrices acting on isospin space. In our work, the starting point is the gap equation, which  within the Hartree-Fock approximation has the form~\cite{ Buballa:2003qv, Klevansky:1992qe}
\begin{equation}
    M=m_{q}-2G\braket{\bar{\psi}\psi},\label{eq:gapgral}
\end{equation}
where $-\braket{\bar{\psi}\psi}$ is the chiral condensate, defined as
\begin{equation}
    -\braket{\bar{\psi}\psi}=\int\frac{d^4p}{(2\pi)^4}{\rm Tr}\left[ iS \left(p\right)\right],\label{eq:condensate}
\end{equation}
and
\begin{equation}
    S(p)=\frac{1}{{\not \! p}-M}
\end{equation}
is the quark propagator with $M$ representing the dynamical quark mass. In the NJL model, $M$ is a constant and the propagator does not exhibit wavefunction renormalization effects.

In order to obtain an explicit form of the gap equation, we need to select an regularization scheme; we adopt a hard cut-off in the spatial momentum. We add the effects of the thermal bath and chemical potential within the Matsubara formalism by doing the replacements~\cite{kapusta_gale_2006}
\begin{equation}
    k \rightarrow (iw_k, \vec{k}),\qquad
    \int{dk_0} f(k_0,\vec k)\rightarrow T
    \sum_{k=-\infty}^{\infty} f(w_k,\vec k).
\end{equation}
Here, $w_k = (2k+1)\pi T+i\mu$ are the fermionic Matsubara frequencies where we already have included a real chemical potential $\mu$. Putting all together, the gap equation has the form
\begin{equation}
    M = m_q+2GT\sum_{k=-\infty}^{\infty}\int_\Lambda\frac{d^3k}{(2\pi)^3}{\rm Tr}\left[iS\left(w_k, \vec{k}\right)\right].\label{gap_int}
\end{equation}
Here, the symbol $\int_\Lambda$ stresses that integrals are regulated.
The evaluation of Eq.~(\ref{gap_int}) can be performed by standard methods of thermal field theory~\cite{kapusta_gale_2006}, leading to 
\begin{equation}
    M = m_q+4N_fN_cG\int_{\Lambda}\frac{d^3k}{(2\pi)^3}\frac{M}{E_k}\bigg(1-n_k(T, \mu)-\bar{n}_k(T, \mu)\bigg)\label{gap_equation},
\end{equation}
where $E_k=\sqrt{\vec{k}^2+M^2}$ is the energy of the quark; $n_k(T,\mu)$ and $\bar{n}_k(T,\mu)$, defined as
\begin{equation}
    n_k(T,\mu)=\frac{1}{e^{(E_k-\mu)/T}+1}, \qquad\bar{n}_k(T,\mu)=\frac{1}{e^{(E_k+\mu)/T}+1},
\end{equation}
are the fermion and antifermion occupation numbers, respectively, and $\Lambda$ is the ultraviolet cut-off.  We  set $N_f = 2$ and $N_c=3$. Finally, in the remaining of the letter, we consider the  chiral limit $m_q\rightarrow0$.

\begin{figure}[t]
    \centering
    \includegraphics[width=\linewidth]{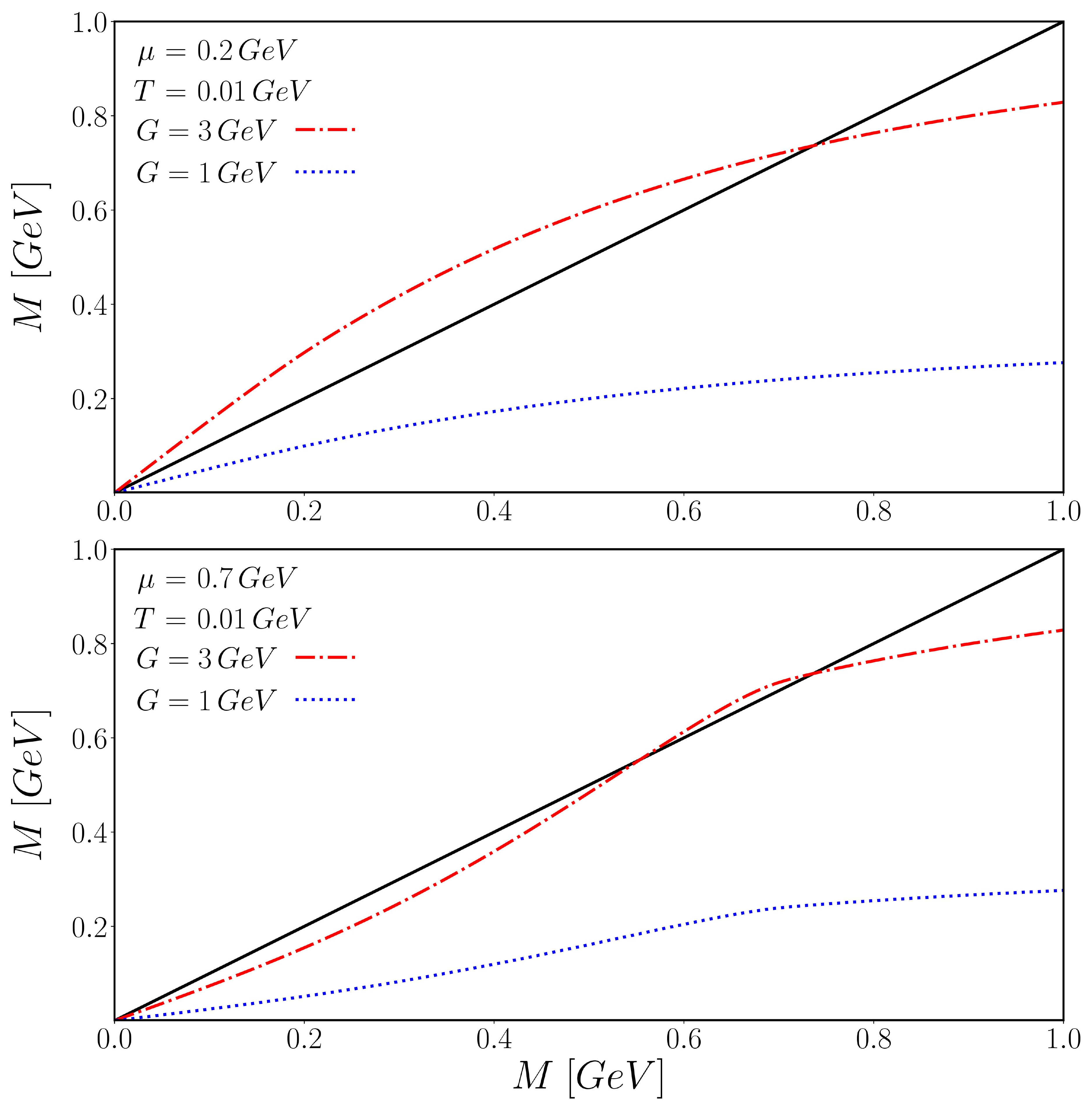}
    \caption{Plot of Eq.~(\ref{gapsplit1}) and Eq.~(\ref{gapsplit2}) for $G=1\, \text{GeV}^{-2}$ (blue, dotted) and $G=3\, \text{GeV}^{-2}$ (red, dashed). For the dotted curves in both panels the value of the coupling is not enough to break chiral symmetry in the model. For stronger values of the coupling (dashed curves), in the upper panel, the chemical potential is still low enough to render the transition of second order. On the other hand, the influence of the very high chemical potential becomes apparent and the transition turns into a first order one. The scale of the graphs is set by $\Lambda = 1$.}
    \label{fig:gap_split}
\end{figure}

In Eq.~(\ref{gap_equation}), the first term, proportional to 1 inside parenthesis,  corresponds exclusively to the vacuum contribution, whereas the terms accompanied by $n_k(T,\mu)$ and $\bar{n}_k(T,\mu)$ are the medium contribution. The first part needs to be regularized. Nevertheless, the cut-off $\Lambda$ in the medium part can be safely set to infinity and the integral is still convergent. On physical grounds, this is in accordance with the idea that the medium is not capable of modifying the ultraviolet properties of the quark propagation. From this observation, we should consider separately the scenarios where the medium is regularized by the cut-off $\Lambda$ or not.

\section{Phase-Diagram and Lagrange multipliers}
\label{S:2}
\begin{figure}[t]
    \centering
    \includegraphics[width=\linewidth]{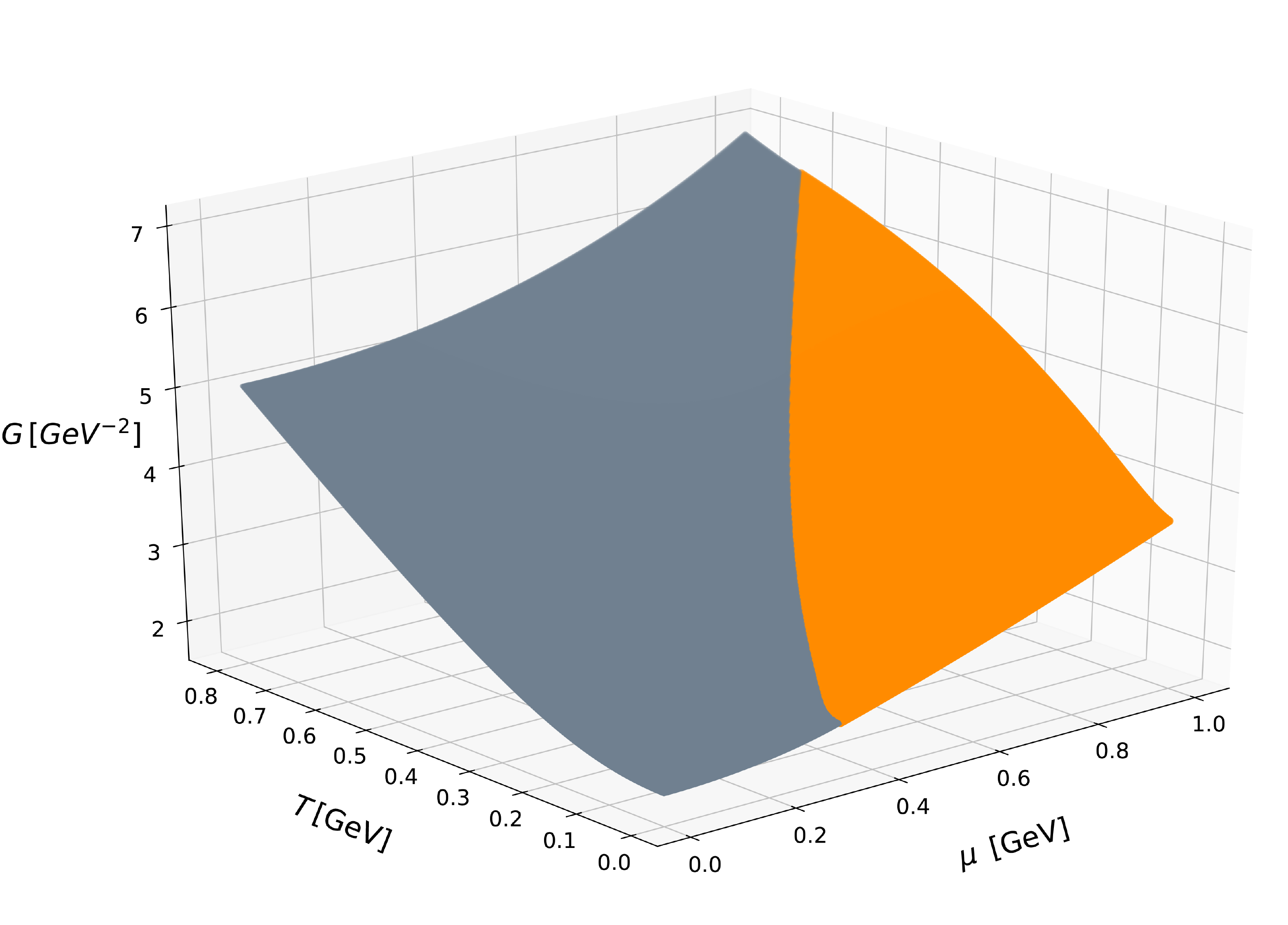}
    \caption{Relationship between the coupling $G$, $\mu$ and $T$ in the phase transition, obtained by solving Eq.~(\ref{gap_equation}) and Eq.~(\ref{critcondition}) simultaneously. In blue, the first order phase, whereas the orange surface represents the second order phase. The phase diagram for a given coupling can be obtained by slicing the surface.}
    \label{fig:diagrama-fase-3d-R}
\end{figure}

Our main interest is to describe the QCD phase-diagram and locate the position of the CEP within the NJL model. To this end, we generalize the method introduced in~\cite{Martinez:2018snm} by adopting the strategy of Lagrange multipliers to the gap equation, considering the coupling constant as such multiplier. This translates to a condition among the parameters $T$ and $\mu$ that define the critical curves that separate the chiral symmetric and asymmetric phases. We start by splitting the gap equation and define left-hand side $H(M)$ and the function $F(M)$ on the right-hand-side as
%
\begin{flalign} 
    H(M)&=M\label{gapsplit1} \\
     F(M)&=4N_fN_c\int_0^{\Lambda}\frac{d^3p}{2\pi^3}\frac{1}{E_p}\bigg(1-n_p(T, \mu)-\bar{n}_p(T, \mu)\bigg),
     \label{gapsplit2}
\end{flalign}
%
such  that the gap equation has the symbolic form
\begin{equation}
    H(M)= G\,M\, F(M).\label{gapeq}
\end{equation}
In Fig. \ref{fig:gap_split} we depict the curves of Eq.~(\ref{gapsplit1}) (black) and Eq.~(\ref{gapsplit2}) (red dashed) as function of the dynamical mass $M$. For the blue (dotted) curves, in both panels, the coupling is not enough to break chiral symmetry in the NJL model and are drawn to lead the eye. Dashed curves correspond to a larger value of the coupling, enough to break chiral symmetry.  In the upper panel,  the small value of the chemical potential $\mu$ (as compared to $\Lambda$, which we set to unity) gives raise to a single positive solution for the dynamical mass (one crossing of the two curves outside the origin). This implies that the transition from the unbroken to the broken chiral symmetry phases is continuous; we observe a second order transition. On the other hand, for the lower panel,  the much larger value of the chemical potential gives raise to a second positive solution and thus the breaking of chiral symmetry under these conditions is discontinuous; here the system undergoes a first order transition. In this scenario, our goal of sketching the phase-diagram reduces to finding the combination of parameters $\mu$, $T$, $\Lambda$ and $G$ that maximize Eq.~(\ref{gapsplit2}) subject to Eq.~(\ref{gapsplit1}). This can be done using the Lagrange Multipliers method, employing the coupling constant $G$ as the Lagrange multiplier.

In this regard, the next step is to equate both the mass gradients of Eqs.~(\ref{gapsplit1}) and~(\ref{gapsplit2}) and include the Lagrange multiplier $G$.  That is
\begin{equation}
    1=G \bigg(F(M)+M\frac{\partial F(M)}{\partial M}\bigg).\label{LM1}
\end{equation}
Upon substituting the gap equation Eq.~(\ref{gap_int}) back into Eq.~(\ref{LM1}),  we get the critical condition
\begin{equation}
    GM^2\frac{\partial F(M)}{\partial M}=0. \label{critcondition}
\end{equation}
The solutions of Eq.~(\ref{critcondition}) describe the value of the mass $M$ for which the slope of the curves in Eqs.~(\ref{gapsplit1}) and~(\ref{gapsplit2}) is the same; the trivial solution $M=0$ correspond to a second order symmetry breaking, whereas an additional positive solution $M>0$ yields a first  order transition for chiral symmetry breaking. If no solution is found, then the system remains in the chirally symmetric phase. Both Eqs.~(\ref{critcondition}) and~Eq.~(\ref{gap_equation}) constitute the information we require to sketch the phase-diagram.

\begin{figure}[t]
    \centering
\includegraphics[width=\linewidth]{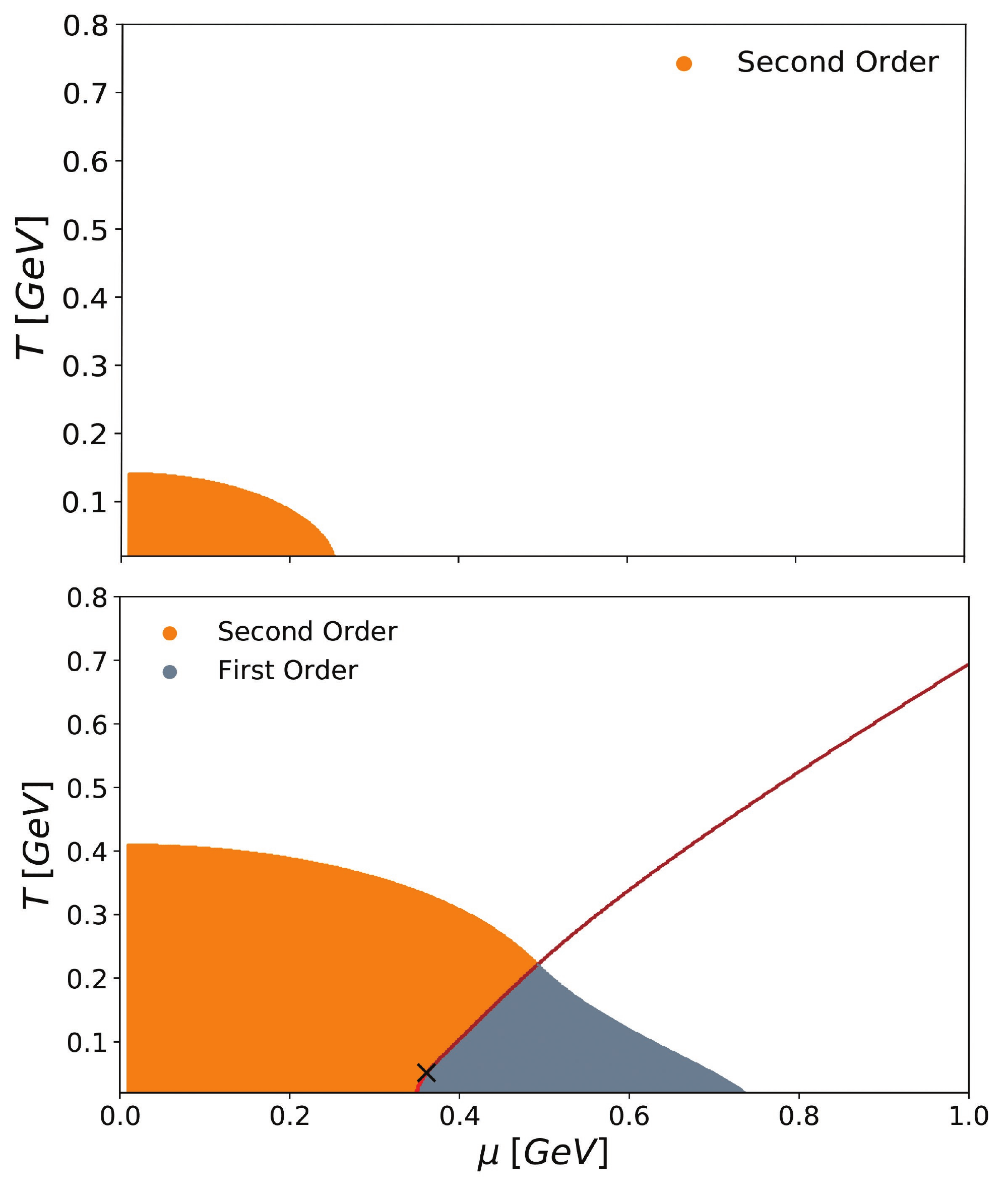}
    \caption{Phase diagram obtained after slicing the surface Fig.~\ref{fig:diagrama-fase-3d-R} in $G=2\, \text{GeV}^{-2}$ (upper panel) and $G=3\, \text{GeV}^{-2}$ lower panel). The orange area depict second order phase, whereas blue area represents first order phase. For the cross mark $G=2.2\,\text{GeV}^-2$. }
    \label{fig:phase-diagram-R}
\end{figure}

The procedure we adopt to this end is the following: We determine tha values of $M$ that solve  Eq.~(\ref{critcondition}) for fixed values of $\mu$ and $T$ and then insert them back into the gap Eq.~(\ref{gap_equation}) to read the value of the coupling $G$ consistent with these values. This leads us to the three dimensional surface that depicts the relation among $\mu$, $T$ and $G$,  Fig.~\ref{fig:diagrama-fase-3d-R}. We fix the regularization parameter $\Lambda=0.942\, \text{GeV}$ in order to reproduce within the model the pion mass at $\mu=0$ and $T=0$, according to Ref.~\cite{Inagaki:2015lma}; we consider the same regulator to all the integrals in the gap equation that so require. To obtain the diagram for a specific value of the coupling, we slice this 3D surface  at the chosen coupling; the projection of this resulting surface is the desired phase-diagram.
In Fig.~\ref{fig:phase-diagram-R} we draw the resulting phase-diagram for $G=2\, \text{GeV}^{-2}$ (upper panel) and $G=3\, \text{GeV}^{-2}$ (lower panel). For $G=2\, \text{GeV}^{-2}$ the only possible transition is a second order one at the boundary of the orange area; here the derivative of the mass function becomes discontinuous at the phase transition and the gap equation has only one positive stable solution. On the other hand, for $G=3\, \text{GeV}^{-2}$, the coupling is strong enough to develop a first order boundary. Here the mass  itself becomes discontinuous at the phase transition depicted in the plot as the boundary of the gray area. We should notice  the appearance of the CEP at $\mu \approx 0.48\, \text{GeV}$  and $T \approx 0.35\, \text{GeV}$ were the three phases meet; the white area being the symmetric phase, the orange one the second order broken phase and gray area, the first order broken phase. The red curve represents the path the position of the CEP changes as the coupling constant becomes stronger. For the sake of illustrating the scale we add a cross-mark that  corresponds to a typical value of the coupling just above the required one to reproduce the pion mass in vacuum $G=2.2\,\text{GeV}^{-2}$.

\begin{figure}[t]
    \centering
    \includegraphics[width=\linewidth]{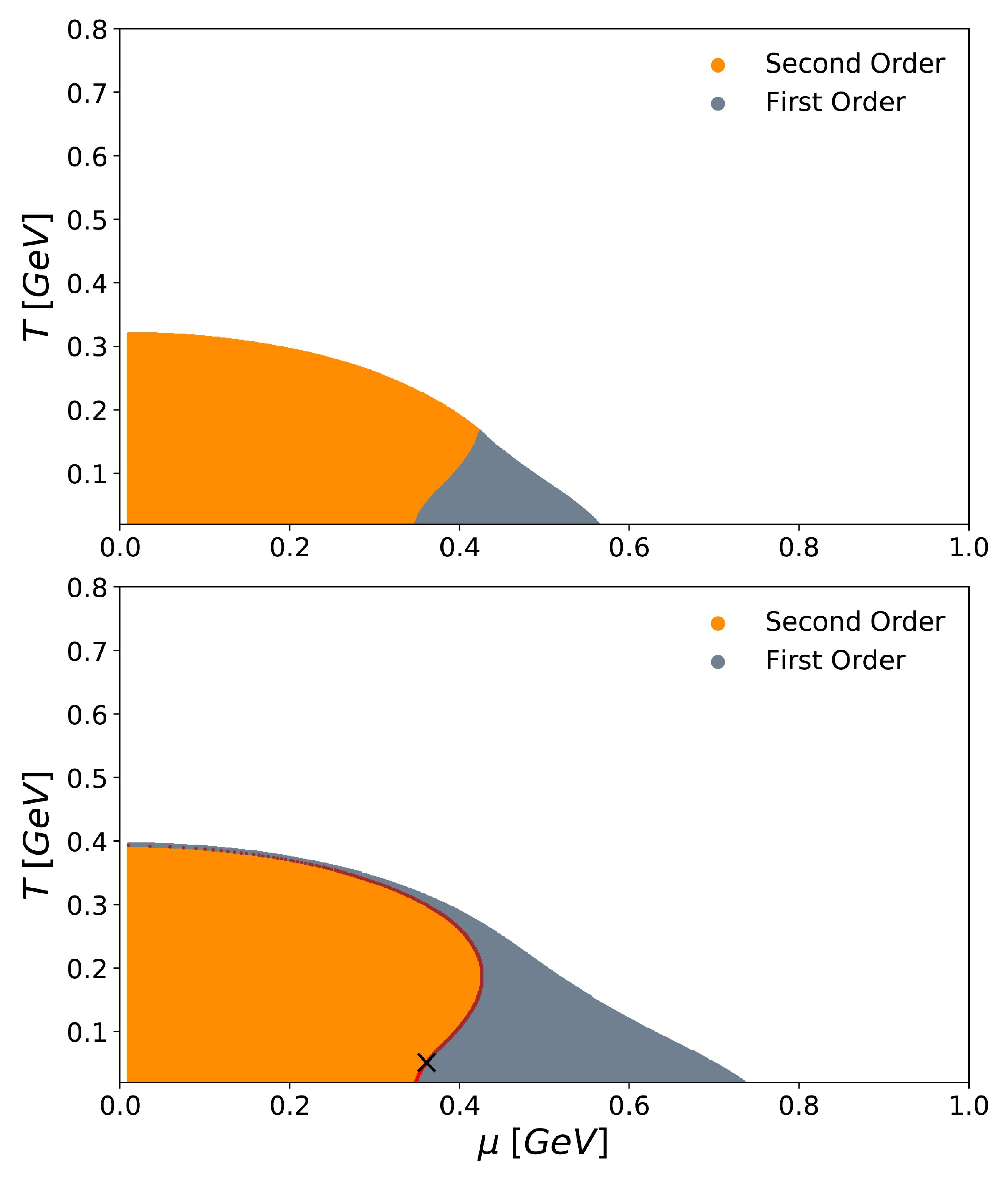}
    \caption{Phase diagram obtained after applying the process described in \ref{S:3} to the gap equation with non regularized medium for $G=2.6\, \text{GeV}^{-2}$ (upper panel) and $G=3\, \text{GeV}^{-2}$ (lower panel). Notice how in the lower panel, $G=3\, \text{GeV}^{-2}$ the first order phase "surrounds" the second order phase. This behaviour is not present in the phase diagram Fig.~\ref{fig:phase-diagram-R}. As in Fig.~\ref{fig:phase-diagram-R} we add a cross mark to illustrate the scale, that is $G=2.2\,\text{GeV}^{-2}$. }
    \label{fig:phase-diagram-NR}
\end{figure}

\begin{figure}[t]
    \centering
    \includegraphics[width=\linewidth]{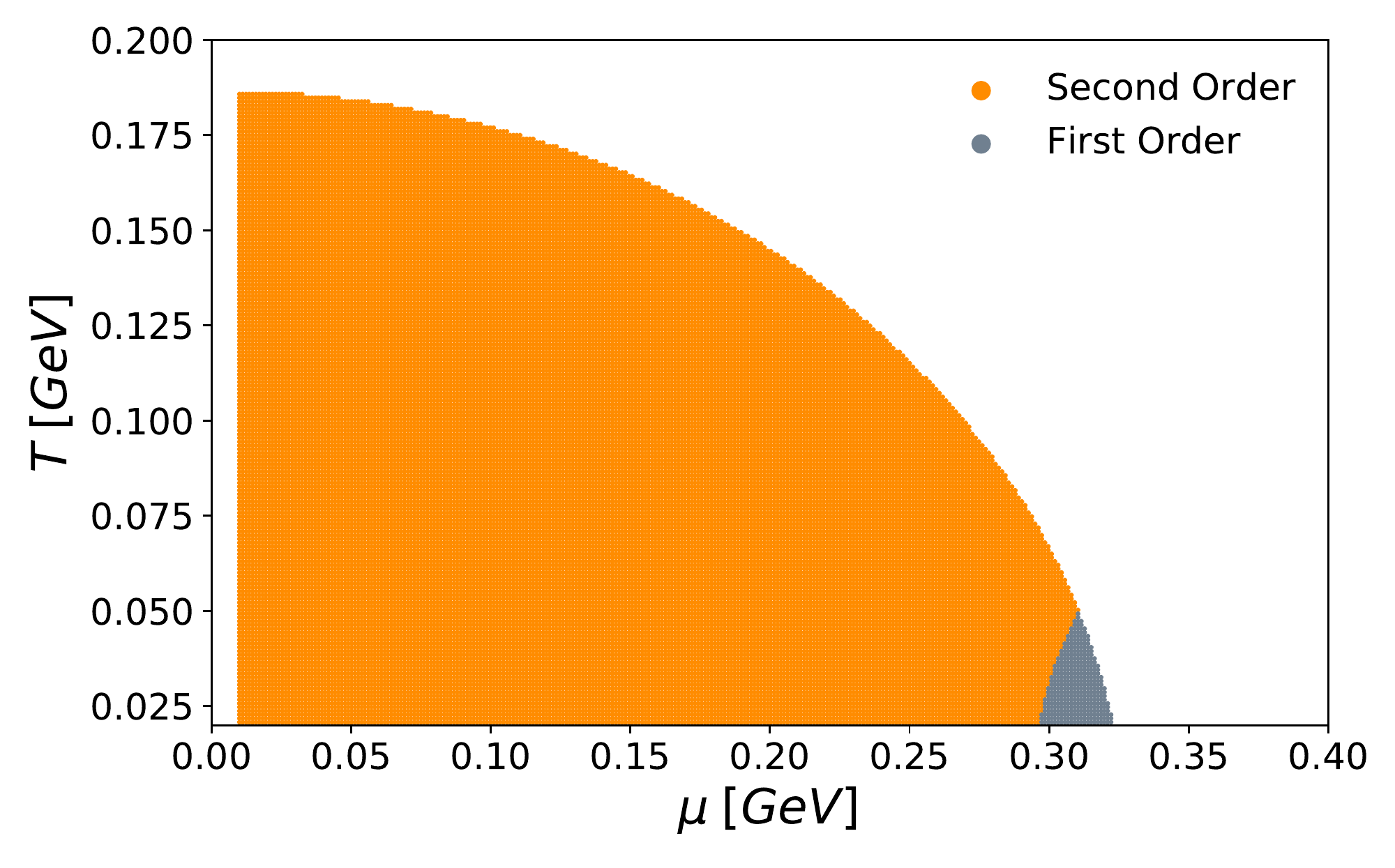}
    \caption{Resulting phase diagram after fixing the critical temperature and chemical potential in Eq.~(\ref{critcondition}) and solving for the coupling and regularization scale. Here $G=3.072\,\text{Gev}^{-2}$ and $\Lambda = 0.8\, \text{GeV}$. The CEP is located in $T=0.049\,\text{GeV}$ and $\mu = 0.31\, \text{GeV}$,  in fair agreement with Ref.~\cite{Ayala:2019skg}. }
    \label{fig:phase-diagram-crit}
\end{figure}

Because our method  makes reference only to  the quark dynamical mass, it can be straightforwardly adapted  to get the phase-diagram for several variations of the gap equation. For example, a comparison can be done directly in the case where the medium is regulated or not. 
With the outlined procedure,  we depict the  phase-diagram for the case where the medium is not regulated in  Fig.~\ref{fig:phase-diagram-NR} for $G=2.6\, \text{GeV}^{-2}$ (upper panel) and $G=3\, \text{GeV}^{-2}$ (lower panel); the phase diagram found was pretty much the same as in Fig.~\ref{fig:phase-diagram-R}, the main difference being the evolution of the first order area as the coupling increases; here, a boundary of first order transition develops such that the second order phase is separated from the symmetric phase 
 covering completely the continuously broken phase at $G=3\, \text{GeV}^{-2}$ and above. We should stress that this behavior is not found in Fig.~\ref{fig:phase-diagram-R} even for large values of the coupling; it is in this regard that the two phases diagram differ, although for strong values of the coupling, where the physical reliability of the model is compromised.
It should be stressed that fixing the parameters of the model is crucial to obtain the phase-diagram with physically meaningful features. Following the procedure outlined in Ref.~\cite{Ayala:2019skg}, by fixing the values of the critical temperature and critical chemical potential found there, we solve the critical condition Eq.~(\ref{critcondition}) for the coupling $G$ and the regularization parameter $\Lambda$. The resulting phase diagram is sketched in Fig.~\ref{fig:phase-diagram-crit}. We observe fair agreemen between phase-diagram obtained from our procedure and the one found in Ref.~\cite{Ayala:2019skg}.

\section{Conclusions and final remarks}~\label{S:3}

In this letter, we have addressed the issue of locating, in the $T\!-\! \mu$-plane, the CEP  for the chiral phase transition within the NJL model. To this end, we have developed an innovative approach to analyze the behavior of the corresponding gap equation in terms of the thermodynamic parameters considering the coupling of the model as a Lagrange multiplier. In this form, we found the critical condition, expressed in Eq.~(\ref{critcondition}), that is capable of yelling apart the various phases of the model, where the dynamical mass or, correspondingly, the chiral condensate either vanishes or varies (continuously and/or discontinuously) with the heat bath parameters.  
In contrast with the common strategy to sketch the phase diagram from the behavior of the chiral condensate or other physical observables -that often leads to a location of the CEP that depends upon the susceptibility under consideration-, 
the critical condition~(\ref{critcondition}) allows us to derive the phase diagram for any value of coupling and what is more, pin-point the location of the CEP without resorting to a specific observable. 

In the simplest local model, it is remarkable that a discontinuous phase transition can only take place when the chemical potential exceeds a critical value even at zero temperature, regardless of the strength of the coupling $G$. Thus the identification of the CEP is uniquely determined from the boundary of the first and second order transitions. In the case when the medium integrals are not regularized, when the coupling is very large, there is a thin strip of first order phase transition that occurs at arbitrarily small values of $\mu$ but large values of the temperature that separates the continuously broken  phase from the symmetric one. This situation has not been resolved in experiments or other theoretical calculations. The explanation to this conundrum can be given by observing that it can only take place for a super-strong value of the coupling, where the physical reliability of the NJL model is jeopardized.

Finally, it is worth mentioning that the applicability of the method presented here to a large set of variations of the NJL and other effective models  where the dynamical mass does not depend on the momentum as well as incorporating other effects like the influence of a magnetic field, a Polyakov loop potential and more. Some of these lines are currently under consideration and findings will be reported somewhere else.

\section*{Acknowledgements}
We acknowledge A. Ayala, A. Bashir and M.~E. Tejeda-Yeomans for insightful discussions. We also acknowledge Consejo Nacional de Ciencia y Tecnolog\'{\i}a (MEXICO) for financial support under grant 256494.





\bibliographystyle{model1-num-names}
\bibliography{sample.bib}







\end{document}